\documentstyle[PASJadd,epsfig,aaspptwo]{PASJ95}
\onecolumn

\def\spose#1{\hbox to 0pt{#1\hss}}
\def\lta{\mathrel{\spose{\lower 3pt\hbox{$\mathchar"218$}}
     \raise 2.0pt\hbox{$\mathchar"13C$}}}
     \def\gta{\mathrel{\spose{\lower 3pt\hbox{$\mathchar"218$}}
          \raise 2.0pt\hbox{$\mathchar"13E$}}}

\begin{document}
\setcounter{page}{001}

\vspace*{3cm}

{\Large\bf Parameters of black holes in sources with periodic 
variability\footnote{To appear in Publ.\ Astron.\ Soc.\ Japan (1999)}}
\vspace*{1cm}

{Old\v{r}ich {\sc{}Semer\'ak} and Vladim\'{\i}r {\sc{}Karas}
\\
{\it{}Faculty of Mathematics and Physics, Charles University Prague,}
\\
{\it{}V~Hole\v{s}ovi\v{c}k\'ach~2, CZ-180\,00~Praha~8, Czech Republic}
\\
{\it{}E-mail: oldrich.semerak@mff.cuni.cz; vladimir.karas@mff.cuni.cz}

and 

Fernando {\sc{}de Felice}
\\
{\it{}Department of Physics ``G.~Galilei'', University of Padova,
Via~Marzolo~8, I-35131~Padova, Italy}
\\
{\it{}E-mail: defelice@padova.infn.it}
}

\vspace*{1cm}

{We discuss a way to deduce parameters of accreting black holes.
The method employs properties of spectral features observed in
radiation from an accretion disk. It is applicable to sources which
exhibit periodic modulation of variability, provided: (i)~Gravitational
field is determined by the black hole and described by the Kerr metric;
(ii)~A thin accretion disk of a negligible mass lies in the equatorial
plane of the hole; (iii)~A secondary object (also with negligible mass)
moves on a slightly inclined almost circular orbit around the black hole
and passes periodically through the disk; (iv)~The collisions result in
observable photometric and spectroscopic features (temporal variability
of the radiation flux and of spectral-line profiles produced in the
disk), which show frequencies of orbital motion and of latitudinal
oscillations; (v)~One can measure width of the spectral line from hot
spots arising in the disk due to collisions with the orbiter, and/or
detect the predicted low-frequency oscillations, induced in the disk.
These frequencies and the line width provide enough information to
determine in physical units three parameters characterizing the source
--- the mass and angular momentum of the central black hole, and radius
of the orbit of the secondary. }


\twocolumn

\section{Introduction}
There is an ever-growing theoretical and observational evidence that the
dark masses present in some galactic nuclei and X-ray binaries are black
holes (Rees 1998). Standard interpretation of observational data employs
a black hole and an accretion disk or a torus (Kato et al.\ 1998). A
number of ways have been suggested of how to deduce the properties of
putative holes. The first rough estimates of their mass were based on
energy considerations and limits implied by the shortest time-scales of
variability. More precise methods became possible with modern
observational techniques like HST, VLBI and X-ray satellites. Our
present paper is relevant for sources with a rotating black hole
surrounded by a thin accretion disk in the equatorial plane and a much
less massive secondary orbiter passing periodically through the disk.
First we briefly recall relevant spectral properties which have been
discussed as possible observational signatures of a black hole (we do
not mention other approaches which involve different physical
mechanisms, e.g.\ gravitational waves).

In galactic nuclei, the existence and the size of the dark central mass
are deduced from the profile of the surface brightness of the nuclei and
from the spatial distribution and dynamics of surrounding gas and stars
(Kormendy \& Richstone 1995). Current optical and radio studies still
probe only scales above $10^{4}$ gravitational radii of the nucleus
which is insufficient to resolve imprints of relativistic effects in the
innermost regions around the centre, in particular to deduce the rate of
its rotation. However, X-ray observations of the fluorescence iron
K$\alpha$ line (Nandra et al.\ 1997), especially its broad and variable
profile skewed to lower energies, indicate that the Doppler and
gravitational redshifts play a role and the emmision takes place very
near a rotating black hole. Several ways have been proposed to estimate
parameters of the black holes in active galactic nuclei which rely on
the source variability: for example, limits on the black-hole mass and
its rate of rotation can be imposed from time delays between variations
of the emission-line strength and of the continuum, and from temporal
changes of the observed emission lines (e.g., Blandford \& McKee 1982;
Stella 1990; Bromley et al.\ 1997).

The evidence for stellar-mass black holes comes from observations of
Galactic X-ray binaries where they are indicated by low luminosity of
one of the components, temporal variability, spectral features (mainly
the presence of relatively strong ultrasoft component and of the hard
X-ray tail), and the minimum mass of the dark component, as estimated
from the mass function (e.g.\ Charles 1997). It has also been argued
(Narayan et al.\ 1997) that the black-hole X-ray binaries could be
distinguished by larger variation in luminosity between the bright and
the faint states than is expected in the sources with neutron stars.
This way of identifying black holes follows from the low radiative
efficiency of advection-dominated accretion flows around black holes.
Present observational evidence offers several objects of this type which
show relatively stable periodic modulation (van der Klis 1997).

In the present contribution it is assumed that the disk remains in the
equatorial plane and the modulating source is represented by a blob
which keeps its identity for several orbital periods. A specific scheme
for the signal modulation can be described in words as follows: The
orbiter intersects the disk periodically and pulls gaseous material out
its plane. This material temporarily obscures the disk, mainly at the
radius of intersection, and affects radiation flux and emission-line
profiles formed in that region and characterized by the observed widths.
Phenomenologically, a spot forms in the disk and then it orbits around
the center, surviving several orbital periods. Radiation of the spot
modulates the observed radiation at the Keplerian frequency which is
different from the frequency of collisions with the disk when the
central black hole rotates. Both ingredients of the model, i.e.\ local
physics of collisions with the disk and observed signal from orbiting
spots, have been described in literature (recently, Ivanov et al.\ 1998;
Karas 1997). In usual terminology the blob represents a hot spot,
however, any perturbation of the disk surface emissivity (either bright,
or less prominent due to obscuration) suffices for the purpose of
phenomenological description.

Subsequent paragraphs describe technical details of the above-mentioned
scenario. We will also show a predicted variable spectral feature 
involving relevant time-scales in a simplified situation.

\section{Assumptions}
Consider a system of a rotating black hole, an equatorial thin disk and
a secondary orbiter (which may be a low-mass black hole --- see Syer et
al.\ 1991; Vokrouhlick\'y \& Karas 1993). If both the disk
and the secondary have negligible masses with respect to the central hole,
the gravitational field is deteremined solely by the central black hole.
In this situation, motion of the secondary is governed by Kerr metric,
which in usual notation (Misner et al.\ 1973) acquires the form
\begin{eqnarray}
{\rm d}s^{2} & = &
  -\Delta\Sigma{\cal A}^{-1}{\rm\,d}t^{2}
   +\Sigma\Delta^{-1}{\rm\,d}r^{2}
   +\Sigma{\rm\,d}\theta^{2}
   \nonumber \\
&& +{\cal A}\Sigma^{-1}\sin^{2}\theta\;
   \left({\rm{d}}\phi-\omega_{\rm K}{\rm{d}}t\right)^{2};
\end{eqnarray}
here Boyer-Lindquist spheroidal coordinates ($t,r,\theta,\phi$) and
geometrized units (in which $c=G=1$, $c$ being the speed of light in
vacuum and $G$ the gravitational constant) have been used, $M$ and $a$
denote mass and specific rotational angular momentum of the centre,
$\Delta=r^{2}-2Mr+a^{2}$,
$\Sigma=r^{2}+a^{2}\cos^{2}\theta$,
${\cal A}=(r^{2}+a^{2})^{2}-{\Delta}a^{2}\sin^{2}\theta$,
and $\omega_{\rm{K}}=2Mar/{\cal A}$.

World-line of the secondary is very close to a geodesic in the Kerr
field, provided that the dynamical effect of passages through the disk
is only weak, that tidal interaction is negligible (the secondary is
assumed much smaller than the typical curvature radius of the field
around), and that gravitational radiation is ignored. The trajectory
undergoes three secular changes due to the interaction with the disk
(Syer et al.\ 1991; Vokrouhlick\'y \& Karas 1993): long-term decrease of
the semi-major axis (spiralling towards the centre due to energy
dissipation in collisions with the disk), circularization (the orbit
becomes spherical, $r={\rm const}$), and gradual tilting of the orbit
into the equatorial plane (the orbit declines to the disk). Since the
time-scale for circularization is shorter than, or of the same order as,
the time-scale necessary to drag the orbit into the disk, one can assume
that the secondary follows a nearly equatorial spherical geodesic at
late stages of evolution of the hole--disk--secondary system. The
secondary remains on this type of orbit for a relatively long time. Let
us note that we ignore gravity of the disk itself; the case of a massive
disk was treated recently by Vokrouhlick\'y \& Karas (1998).

The resulting system is characterized by two angular frequencies ---
that of the azimuthal revolution of the orbiter around a Kerr black
hole, $\omega$, and that of its latitudinal oscillations about the
equatorial plane, $\Omega_{\infty}$. Both frequencies are in principle
measurable at infinity provided that the passages of the orbiter produce
strong enough modulation of the disk radiation (Karas \& Vokrouhlick\'y
1994 illustrated, by Fourier analysis of simulated photometric data, how
these two peaks can be recognized in the power spectrum).

Relevant parameters of a quasi-circular orbit precessing with low
amplitude about the equatorial plane of the Kerr spacetime are given in
next section. Relations derived in Sect.\ \ref{eqset} provide the
parameters of the central black hole in terms of four observables ---
the three frequencies ($\Omega_\infty$, $\omega_+$, and $\kappa$, as
they are introduced in the text) and a spectral-line width. Equations
can be solved easily if the secondary is not too close to the black
hole (cf.\ the Appendix).

\section{Parameters of the Precessing Orbit}
For a general bound geodesic in the Kerr spacetime, the frequencies of
the azimuthal and latitudinal motion can be given in terms of elliptic
integrals (Karas \& Vokrouhlick\'y 1994). Relevant expressions acquire a
simpler form in case of spherical geodesics (Wilkins 1972) and simplify
still further for almost equatorial orbits. The azimuthal angular
velocity $\omega={\rm d}\phi/{\rm d}t$ (with respect to a distant
observer at rest) is then approximated by the Keplerian circular
frequency
\begin{equation}  \label{omega}
  \omega_{\pm}=(a+1/y_{\pm})^{-1},
\end{equation}
where $y_{\pm}=y(\omega_{\pm})=\pm\sqrt{M/r^{3}}$ and the
upper/lower sign corresponds to the prograde/retrograde
orbit.
We keep both cases for completeness, but only prograde trajectories
(plus sign) will be considered later (the accretion disk is more likely
to be corotating with the central black hole and the interaction with
the disk makes also the secondary corotate eventually).

The proper angular frequency $|\Omega|$ of small latitudinal
harmonic oscillations about the equatorial plane is given, for a
spherical orbit with steady radial component of acceleration, by (de
Felice \& Usseglio-Tomasset 1996; Semer\'ak \& de Felice 1997)
\begin{equation}  \label{Omega}
 \Omega^{2}=
 (u^{t}/r)^{2}
 \left\{\Delta\omega^{2}+
        2y_{\pm}^{2}[a-(r^{2}+a^{2})\,\omega]^{2}\right\}.
\end{equation}
Here $\omega={\rm{}const}$, and the time component of
four-velocity, given by
$(u^{t})^{-2}=-g_{tt}-2g_{t\phi}\omega-g_{\phi\phi}\omega^{2}$,
is approximated by the equatorial value
\begin{equation}  \label{ut}
  (u^{t})^{-2}=1-(2M/r)(1-a\omega)^{2}-(r^{2}+a^{2})\,\omega^{2}.
\end{equation}
Frequency of the collisions with the disk is twice the proper
frequency $|\Omega|$; the corresponding value measured by a
distant observer is $|\Omega_{\infty}|=|\Omega|/u^{t}$.
From eqs.\ (\ref{omega})--(\ref{Omega}), the explicit expression
for a {\em free\/} orbit is
\begin{eqnarray}  \label{Omegaobs}
  \Omega_{\infty}^{2}
  &=&\omega_{\pm}^{2}\,(1-4ay_{\pm}+3a^{2}/r^{2}) \nonumber \\
  &=&\frac{\omega_{\pm}^{2}}{r^{2}}
     \left[\Delta+2(\sqrt{Mr}\mp a)^{2}\right].
\end{eqnarray}
We remind that the frequency of latitudinal oscillations stands in the
dispersion relation for corrugation waves in accretion disks. It has
been suggested that $\Omega_{\infty}$ could be detected in some QPO
sources (Kato 1990).

The two eqs.\ (\ref{omega}) and (\ref{Omegaobs}) are not enough to
fix the three unknowns $M$, $a$ and $r$. Another piece of
information can be provided by the width of the emission line
which is formed and modulated at the radius of successive
collisions between the secondary and the disk. A stationary disk
produces the well-known double-horn line profile which is more or
less pronounced according to the inclination angle of the source.
Radiation from each element in the disk experiences a frequency
shift (Fanton et al.\ 1997)
\begin{equation}  \label{g}
 \!g\equiv\frac{\nu_{\rm{}obs}}{\nu_{\rm{}em}}=
 u^{t}\,\frac{\omega_{\pm}}{y_{\pm}}
 \left[1-\frac{2M}{r}+y_{\pm}
       (a+\sqrt{\Delta}\,e_{\hat{\phi}})\right]
\end{equation}
where $e_{\hat\phi}$ is an emission direction cosinus (an
azimuthal component of the unit vector along the direction of
emission of a given photon, measured in the emitter's local frame),
and
\begin{eqnarray}
 (u^{t})^{2}&=&\frac{y_{\pm}^{2}}{\omega_{\pm}^{2}}\,
               \frac{1}{1-3M/r+2ay_{\pm}} \nonumber \\
            &=&\frac{M}{r}\,\frac{1}{\omega_{\pm}^{2}}\,
               \frac{1}{\Delta-(\sqrt{Mr}\mp a)^{2}}
\end{eqnarray}
follows from (\ref{ut}) and (\ref{omega}).
The observed total width of the line arises from the different
frequency shifts $g$ carried by photons which reach the observer
at infinity. Since $g$ depends on the emission cosinus
$e_{\hat{\phi}}$, the maximum width of the line $\delta g$ (which is
a result of integration over one entire circle) corresponds to the
range of $e_{\hat{\phi}}$ consistent with the escape to infinity,
$\delta e_{\hat{\phi}}$. We have from eq.\ (\ref{g})
\begin{equation}
 (\delta{g})=(\delta{}e_{\hat{\phi}})\,
 u^{t}\omega_{\pm}\sqrt{\Delta},
\end{equation}
and thus
\begin{eqnarray}  \label{delta}
 \delta^{2}\equiv
 \frac{(\delta g)^{2}}{(\delta e_{\hat{\phi}})^{2}}
 &=&\frac{M}{r}\,\frac{1-2M/r+a^{2}/r^{2}}{1-3M/r+2ay_{\pm}}
     \nonumber \\
 &=&\frac{M}{r}\,\frac{\Delta}{\Delta-(\sqrt{Mr}\mp a)^{2}} \; .
\end{eqnarray}

From this equation it appears convenient to include the parameter
$\delta$ in our consideration because it can be written in the form
independent of inclination angle. To determine $\delta$ in terms of
$\delta{g}$, one needs to fix the range of $\delta{}e_{\hat{\phi}}$.
This can be easily estimated for $r\gg M$ (cf. Appendix). In general,
$\delta{}e_{\hat{\phi}}$ is at most 2
(which is acquired for $\theta_{\rm{o}}=\pi/2$, i.e.\ edge-on view of
the disk), and it decreases with the inclination angle $\theta_{\rm{o}}$
of the black hole--disk system relative to the line of sight (it also
depends, quite weakly, on the rotational parameter $a$ and the radius of
emission $r$). It appears unpractical to deal with a lengthy analytic
expression for $\delta{}e_{\hat{\phi}}$. Instead, given $a/M$, $r/M$ and
$\theta_{\rm{o}}$, the actual value of $\delta{}e_{\hat{\phi}}$ can be
found by numerical ray tracing (Fanton et al.\ 1997).

We simulated a time-variable line profile to illustrate the main
features expected in the observed signal (we used a modification of the
code described by Karas and Vokrouhlick\'y 1994). Figure~1 shows the
measured count rate in the line (background subtracted) as a function of
energy and orbital phase in the equatorial plane of an extreme ($a=M$)
Kerr  black hole. In this example, only that contribution to the total
light is taken into account which originates at $10M\lta{}r\lta{}14M$
where an orbiter affects the disk. Energy is normalized to the emission
energy of the line in the local frame of the disk material, as usual
(energy axis thus indicates the redshift factor $g$). Phase axis is
normalized with respect to Keplerian orbital motion at the central
radius $r=12M$, and interval of two periods is shown for clarity. One
can observe:

\noindent{}
(i)~Modulation of the count rate in the line by the orbiter crossing the
disk upwards at radius $r$ with period $T\equiv2\pi/\Omega_\infty(r)$.
A flare is triggered at the moment of crossing and then falls off
with time as the orbiting spot decays and the material swept out of the
disk cools down (characteristic time-scale for this decay is taken to be
two orbital periods; it is a phenomenological parameter here).

\noindent{}
(ii)~Another modulation by the orbiting spot with period
$t\equiv2\pi/\omega_+$. Corresponding maxima of the count rate can be
seen at phase equal to 0.5, 1.5, \ldots\ (due to Doppler boosting).

\noindent{}
(iii)~Orbital variation $\delta{g}$ of the redshift factor (also with
periodicity $t$) in projection down into the energy-phase plane.

In Fig.~1 inclination angle has been taken 40 deg. Obviously, any
non-zero inclination would lead to a qualitatively similar picture.
However, at higher inclinations ($\gta60$\,deg) another local maximum
would develop in the observed flux due to the lensing effect. (Its
position is advancing in phase by $\approx0.25$ with respect to the
Doppler peak.) Difference in the two frequencies, $\Omega_\infty$ and
$\omega_+$, discussed in previous paragraphs translates to the time gap
$T-t$. Let us emphasize that this figure offers a qualitative
illustration of the predicted time-variable spectral feature. In
a more realistic model one needs to take into account a proper
method of background continuum subtraction (underlying continuum has
been ignored here). Mutual interplay between spectral features (around
6.4 keV) and continuum has been subject of recent works (Young et al.\
1998; $\dot{\rm{Z}}$ycki et al.\ 1998; Martocchia et al.\ 1999) which
show substantial smoothing of the profile in some situations.

\begin{Fv}{1}
{0pc
\epsfxsize=\hsize\epsfbox{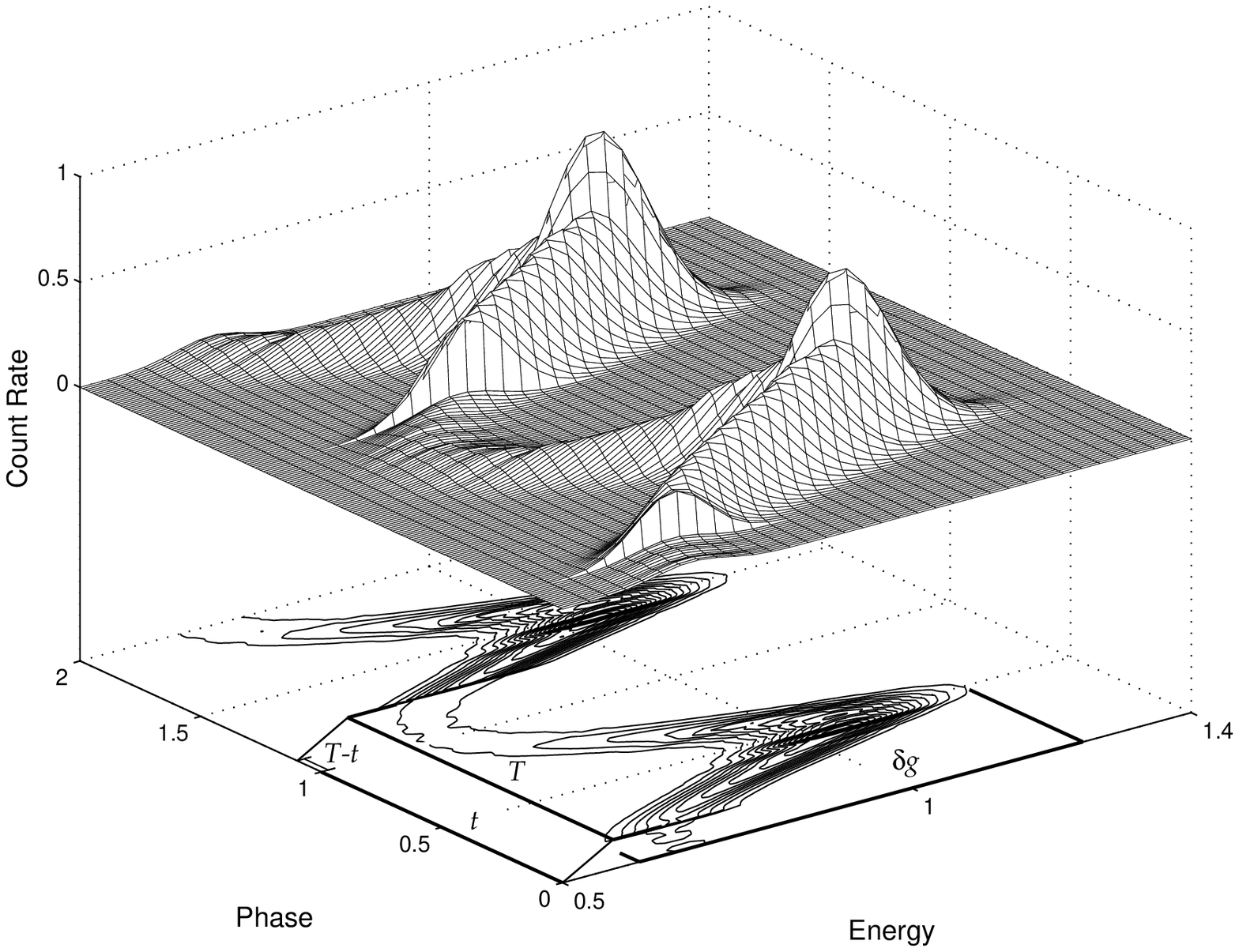}
}
{\protect\small{}
Simulation of the time-variable line profile from a hot spot orbiting with
the disk. A flare occurs at the moment of the star-disk collision (the sharp
increase in flux), giving rise to a spot on the disc surface. Observed
count rate is then modified by orbital motion of the spot (period $t$)
and by relativistic lensing (with a corresponding change of energy in
the range $\delta{g}$). The spot decays gently with time, until another
one is generated in a subsequent collision (i.e.\ after period $T$).
This graph has been obtained by ray tracing in the Kerr metric. Decay
time of the spot is taken as a free parameter. See the text for details.}
\end{Fv}

Uncertainties connected with the badly known $\delta{}e_{\hat{\phi}}$ can
be overcome by introducing another independent observable. A promising
possibility arises from a more realistic view of the star-disk
collision. As mentioned above, the collisions disturb the disk
periodically with latitudinal frequency (\ref{Omegaobs}). If the orbiter
passes across the disk at small $r$, it may induce waves in the disk
matter which remain trapped in the range of several gravitational radii
and are governed by epicyclic frequency $\kappa$ (Okazaki et al.\ 1987;
Nowak \& Wagoner 1992):
\begin{eqnarray}  \label{kappa}
  \kappa^{2}
  &=&\omega_{\pm}^{2}\,
     (1-6M/r+8ay_{\pm}-3a^{2}/r^{2}) \nonumber \\
  &=&\frac{\omega_{\pm}^{2}}{r^{2}}
     \left[\Delta-4(\sqrt{Mr}\mp a)^{2}\right].
\end{eqnarray}
These modes can be excited and trapped in a rather narrow range of
radius, and one can thus expect that they will reflect just the orbital
radius of the satellite. However, the problem of distinguishing between
different possible modes is a subject of much debate and has not yet
been settled.

\section{Determining {\protect\boldmath{$M$}}, {\protect\boldmath{$a$}}, 
         and {\protect\boldmath{$r$}}}
\label{eqset}
The formulas (\ref{omega}), (\ref{Omegaobs}), (\ref{delta}) and/or
(\ref{kappa}) provide a closed system of ordinary equations for $M$,
$a$ and $r$ (and possibly $\delta{}e_{\hat{\phi}}$, if one succeeds in
measuring all four observables) as functions of the quantities
$\omega_{\pm}$, $|\Omega_{\infty}|$, $|\delta|$ and/or $|\kappa|$. The
equations can be tackled in different manners depending on which of the
quantities is determined with the best confidence. Now we discuss
derivation of the most desired combinations of parameters, $r/M$ and
$a/M$, from different starting relations:

\noindent{}
(i)~With the knowledge of $|\Omega_{\infty}|$ and $|\kappa|$
one can start from eqs.\ (\ref{Omegaobs}) and (\ref{kappa}) and find
\begin{eqnarray}
  \frac{\Delta}{r^{2}}&=&
  \frac{2\Omega_{\infty}^{2}+\kappa^{2}}{3\omega_{\pm}^{2}}, \\
  \frac{(\sqrt{Mr}\mp a)^{2}}{r^{2}}&=&
  \frac{\Omega_{\infty}^{2}-\kappa^{2}}{6\omega_{\pm}^{2}},
\end{eqnarray}
which can be solved with respect to $M/r$ and $a/r$:
\begin{eqnarray}
  \frac{M}{r}&=&
  1-\frac{\Omega_{\infty}^{2}+5\kappa^{2}}{6\omega_{\pm}^{2}}
 -2\left(\frac{\Omega_{\infty}^{2}-\kappa^{2}}
              {6\omega_{\pm}^{2}}\right)^{\!\!1/2}
  \nonumber \\ & &
   \times
   \left(1-\frac{\Omega_{\infty}^{2}+2\kappa^{2}}
                {3\omega_{\pm}^{2}}\right)^{\!\!1/2}\!\!\!,~~~
  \label{M/r} \\
  \frac{a^{2}}{r^{2}}&=&
  \frac{2\Omega_{\infty}^{2}+\kappa^{2}}{3\omega_{\pm}^{2}}
  +\frac{2M}{r}-1. \label{a/r}
\end{eqnarray}
The values of $r/M$ and $a/M$ are reached then by obvious
manipulations. One more relation is needed to obtain the absolute
values $M$, $a$ and $r$. Using eq.\ (\ref{omega}), we fix
\begin{equation}  \label{1/r}
  \frac{1}{r}=
  \left(\frac{a}{r}\pm\sqrt{\frac{r}{M}}\right)\omega_{\pm};
\end{equation}
$M$ and $a$ follow from above.

\noindent{}
(ii)~Starting from eq.\ (\ref{delta}), a convenient possibility is
to combine it with eqs.\ (\ref{Omegaobs}) and (\ref{kappa}). We
find
\begin{equation}
  \delta^{2}=\frac{2M}{3r}
             \frac{2\Omega_{\infty}^{2}+\kappa^{2}}
                  {\Omega_{\infty}^{2}+\kappa^{2}},
\end{equation}
and thus the value of $M/r$ follows immediately. This can be compared
with the result (\ref{M/r}) to fix (or verify) the value of
$\delta{e_{\hat{\phi}}}$.

\noindent{}
(iii)~Suppose we know $\omega_{\pm}$, $|\Omega_{\infty}|$ and
$|\delta|$. Introducing
\begin{equation}
  a=\omega_{\pm}^{-1}-y_{\pm}^{-1}
\end{equation}
from eq.\ (\ref{omega}) to eqs.\ (\ref{Omegaobs}) and
(\ref{delta}), we obtain
\begin{equation}
  r^{2}=\frac{3\,(1-\omega_{\pm}/y_{\pm})^{2}}
             {4\omega_{\pm}y_{\pm}+
              \Omega_{\infty}^{2}-
              5\omega_{\pm}^{2}};
\end{equation}
$y_{\pm}$ is then given by quartic equation. This is rather
cumbersome to be handled analytically and it is more suitable to
find the solution numerically for each given set of data.
However, a simple explicit solution can be found analytically if
$r^{2}{\gg}a^{2}$; see the Appendix.

\noindent{}
(iv)~The case when one measures $\omega_{\pm}$, $|\kappa|$ and
$|\delta|$, and thus turns to eqs.\ (\ref{omega}), (\ref{kappa})
and (\ref{delta}), also leads to a quartic equation.

In the Schwarzschild case, $a=0$, eqs.\ (\ref{omega}) and
(\ref{Omegaobs}) are reduced to
\begin{equation}  \label{oy,Schw}
  \omega_{\pm}=y_{\pm}=\pm|\Omega_{\infty}|=
  \pm\sqrt{M/r^{3}},
\end{equation}
eq.\ (\ref{delta}) reads
\begin{equation}  \label{delta,Schw}
  \delta^{2}=r^{2}\omega_{\pm}^{2}\,
             \frac{1-2r^{2}\omega_{\pm}^{2}}
                  {1-3r^{2}\omega_{\pm}^{2}}
\end{equation}
and eq.\ (\ref{kappa})
\begin{equation}  \label{kappa,Schw}
  \kappa^{2}=\omega_{\pm}^{2}(1-6M/r).
\end{equation}
The physical solution of eq.\ (\ref{delta,Schw}) is
\begin{equation}
\!  r^{2}=(4\omega_{\pm}^{2})^{-1}
        \left[3\delta^{2}+1
              -\sqrt{(3\delta^{2}+1)^{2}-8\delta^{2}}\right]
\end{equation}
while that of eq.\ (\ref{kappa,Schw}) 
\begin{equation}
  r^{2}=(6\omega_{\pm}^{2})^{-1}
        \left(1-\kappa^{2}/\omega_{\pm}^{2}\right);
\end{equation}
$M$ follows then from eq.\ (\ref{oy,Schw}).

Note that the above formulas are written in geometrized units.
Corresponding quantities in physical units are obtained by the
following conversions:
\begin{eqnarray}
  \frac{M^{\rm phys}}{M^{\rm phys}_{\odot}}&=&
     \frac{M}{1.477\times 10^{5}{\rm cm}} \, , \\
  a^{\rm phys}=ca, \;&\;&\; r^{\rm phys}=r;
\end{eqnarray}
also
\begin{equation}
  \frac{a}{M}=\frac{a^{\rm phys}}{GM^{\rm phys}/c} \,,\quad
  \frac{r}{M}=\frac{r^{\rm phys}}{GM^{\rm phys}/c^{2}} \,.
\end{equation}
To obtain frequency in physical units [Hz] (either from $\omega$
or $\Omega$), one uses the relation
\begin{equation}
\!  f^{\rm phys}=\frac{\omega^{\rm phys}}{2\pi}
              =\frac{c\omega}{2\pi}
              =(4.77\times 10^{9}{\rm cm/s})\omega.
\end{equation}
In usual notation of relativistic astrophysics the
(geometrized) frequencies are scaled by $M^{-1}$ (as in
Tab.~1). Their numerical values must be multiplied by factor
\begin{equation}
  \frac{c}{2\pi M}=(3.231\times 10^{4})\,
                \left(\frac{M}{M_{\odot}}\right)^{\!-1}\,[{\rm{Hz}}].
\end{equation}

\section{Discussion}
\label{discus} We examined an interesting possibility of determining the
intrinsic parameters of a black hole -- accretion disk system in sources
with periodic modulation of variability which is caused by an orbiting
satellite. The method combines different pieces of information contained
in time-variable spectral and photometric signal, which have been
discussed (widely but separately) in the recent literature. It can
provide all the relevant quantities in physical units, but of course it
can also be combined with another independent determination of some of
the parameters. We assumed that the orbiting secondary is only weakly
affected by the disk, which allowed us to approximate its orbit by a
nearly equatorial spherical geodesic. This restriction must be abandoned
if the secondary is not compact enough.

The proposed approach can be applied to galactic nuclei as well as to
stellar-mass objects in the Galaxy, but identity and physical properties
of the secondary depend on the mass of the centre around which it
revolves. In the latter case the lifetime of the system is more
restricted by tidal interactions and the secondary appears unlikely to
survive long enough. Effects of tidal distortion of a stellar body
passing near a black hole have been discussed by several authors who
show that the star becomes squeezed and heated (Rees 1988; Luminet \&
Marck 1985; Evans \& Kochanek 1989). Tidal disruption occurs in case of
a close encounter, when the satellite plunges below the tidal radius
$R_{\rm{t}}\sim10^{11}\left(M/M_*\right)^{1/3}\left(R_*/R_\odot\right)$
cm, where $M_*$ and $R_*$ denote the mass and the radius of the
satellite. On the other hand, distortions of a compact satellite near a
supermassive black hole are negligible. Duration of a spot or a vortex
in an accretion disk has been also discussed by several authors but
remains still uncertain (Adams \& Watkins 1995; Bracco et al.\ 1998). We
assumed that the lifetime exceeds the corresponding orbital period.

Several possible targets have been discussed already. For example,
optical outbursts in the blazar OJ~287 have recently been modelled in
terms of a black-hole binary system by Villata et al.\ (1998). The
source exhibits several time-scales: feature-less short-term
variability, 12-yr cycle, and, possibly, a 60-yr cycle. These authors,
however, presume both components to be of comparable masses while in our
calculation frequencies are determined under the assumption that the
secondary is much less massive than the primary (cf.\ also Sundelius et
al.\ 1997). There are more BL Lac objects which show a periodic
component of optical variability but time-scales are always
$\gta10\,$yrs in the optical band, and statistical compilation of data
is therefore very difficult (Liu et al.\ 1997). As a better example, a
16-hr periodicity was reported in the X-ray signal from the Seyfert
galaxy IRAS 18325-5926 (Iwasawa et al.\ 1998).

We conclude with M.~J.\ Rees (1998): {\sl{There is a real chance that
someday observers will find evidence that an AGN is being modulated by
an orbiting star, which would act as a test particle whose orbital
precession would probe the metric in the domain where the distinctive
features of the Kerr geometry should show up clearly.}} Indeed, to be
able to resolve the effects of general relativity in AGN, the satellite
would have to orbit near the centre and produce periodicity on a
few-hours time-scale.

\def\lb#1{{\protect\linebreak[#1]}}
\par
\vspace{1pc}
\par

O.\,S.\ thanks the University of Padova and the International Centre for
Theoretical Physics in Trieste for hospitality, and he acknowledges
support from the grant GACR 202/\lb{2}99/\lb{2}0261 in Prague.
F.\,de\,F.\ thanks for support from Agenzia Spaziale Italiana,
Gruppo Nazionale per la Fisica Matematica del C.N.R., and Ministero
della Ricerca Scientifica e Tecnologica of Italy. V.\,K.\ thanks for
hospitality of the International School for Advanced Studies in Trieste;
support from the grants GACR 205/\lb{2}97/\lb{2}1165 and
202/\lb{2}98/\lb{2}0522 is acknowledged.

\section{Appendix. A Simple Explicit Solution of Eqs.\ (\protect\ref{omega}),
 (\protect\ref{Omegaobs}) and (\protect\ref{delta}) for
 {\protect\boldmath{$r^{2}{\protect\gg}a^{2}$}}}
\label{simple}

\begin{table}[t]
\begin{center}
\small
\begin{verse}
Table~1.\hspace{4pt}Accuracy of approximative relations
 (\protect\ref{r})--(\protect\ref{a}). Here, $a$ and $r$ stand as input
 parameters of the system (black-hole angular momentum and radius of the
 orbit of the secondary, respectively). Corresponding values of
 $\protect\tilde{a}$ and $\protect\tilde{r}$ are deduced from our
 approximative equations as given in the Appendix.\\
\end{verse}
\vspace{6pt}
\begin{tabular}{c@{~~}c@{~~}c@{~~}c@{~~}c@{~~}c@{~~}c}
\hline
$a$ & $r$ & $\omega_+$ & $\Omega_\infty$ & $\delta$ &
$\tilde{a}$ & $\tilde{r}$ \\ \hline\hline
0.30 & ~5 & 0.08711 & 0.08279 & 0.5158 & 0.335 & ~5.4 \\
     & 10 & 0.03133 & 0.03077 & 0.3338 & 0.311 & 10.2 \\
     & 20 & 0.01114 & 0.01107 & 0.2292 & 0.303 & 20.1 \\
     & 40 & 0.00395 & 0.00394 & 0.1600 & 0.301 & 40.1 \\
0.60 & ~5 & 0.08489 & 0.07727 & 0.4921 & 0.762 & ~5.7 \\
     & 10 & 0.03103 & 0.03001 & 0.3300 & 0.644 & 10.4 \\
     & 20 & 0.01111 & 0.01097 & 0.2284 & 0.614 & 20.3 \\
     & 40 & 0.00394 & 0.00393 & 0.1598 & 0.605 & 40.2 \\
0.85 & ~5 & 0.08312 & 0.07353 & 0.4773 & 1.305 & ~5.9 \\
     & 10 & 0.03080 & 0.02944 & 0.3273 & 0.943 & 10.6 \\
     & 20 & 0.01108 & 0.01089 & 0.2278 & 0.878 & 20.4 \\
     & 40 & 0.00394 & 0.00392 & 0.1597 & 0.859 & 40.3 \\
1.00 & ~5 & 0.08210 & 0.07168 & 0.4702 & 1.636 & ~6.0 \\
     & 10 & 0.03065 & 0.02914 & 0.3258 & 1.135 & 10.7 \\
     & 20 & 0.01106 & 0.01085 & 0.2274 & 1.039 & 20.4 \\
     & 40 & 0.00394 & 0.00391 & 0.1596 & 1.013 & 40.3 \\
\hline
\end{tabular}
\end{center}
\end{table}

It is relevant to suppose that the secondary orbits at $r\gta10M$, and
typically at $r\sim30M$. In this region a significant part of the disk
radiation arises. Also, the secondary would be tidally disrupted if it
plunges too close to the black hole horizon (to $r\sim M$), while at too
large orbital radius the relativistic dragging effect vanishes (with
$\propto r^{-3}$) and the frequencies $\omega_{\pm}$,
$|\Omega_{\infty}|$ would become indistinguishable.

At large enough radii ($r\sim30M$), several terms in eqs.\
(\ref{omega}), (\ref{Omegaobs}) and (\ref{delta}) become negligible when
compared with the rest. Ignoring them, one arrives at a simplified set
of relations which keeps an acceptable precision of the resulting
parameters without the need to solve the fourth-order equation. (Exact
equations can be used to improve the estimates numerically.) One
chooses some particular approximation according to what precision is
desired at each of the quantities $M$, $a$, $r$, and also according to
precision of each of the input data $|\omega_{\pm}|$,
$|\Omega_{\infty}|$, $|\delta|$. For instance, since $ay_{\pm}<1/30$, we
can take
\[
\omega_{\pm}\simeq{}y_{\pm}(1-ay_{\pm}),\;\;\;
y_{\pm}\simeq\omega_{\pm}(1+a\omega_{\pm})
\]
(for the determination of $M$ from the known $r$, or vice versa,
it is even sufficient to neglect $ay_{\pm}$ altogether and use
just the Schwarzschild form of equations). Other two relevant bounds are:
\[
\frac{a^{2}/r^{2}}{2M/r}<\frac{1}{20},\quad
\frac{a^{2}/r^{2}}{ay_{\pm}}=\frac{ay_{\pm}}{M/r}=
 \frac{a}{\sqrt{Mr}}<\frac{1}{3}.
\]

An example of a reasonable estimate is provided as follows (we suppose a
prograde orbit of the secondary, and consequently we take plus sign in
the relations): Restricting to the Schwarzschild limit of eq.\
(\ref{delta}),
\begin{equation}
 \delta^{2}=\frac{M}{r}\,\frac{1-2M/r}{1-3M/r},
\end{equation}
one comes to an approximative value for $M/r$
(the approximative values will be denoted by a tilde below),
\begin{equation}  \label{r}
 4\tilde{M}/\tilde{r}=
 1+3\delta^{2}
       -\sqrt{(1+3\delta^{2})^{2}-8\delta^{2}}.
\end{equation}
Approximative expression for $a/r$ follows then from eq.~(\ref{Omegaobs}):
\begin{equation}
 \frac{3\tilde{a}}{\tilde{r}}=
 2\sqrt{\frac{\tilde{M}}{\tilde{r}}}-
 \sqrt{4\frac{\tilde{M}}{\tilde{r}}-
 3\left(1-\frac{\Omega_{\infty}^{2}}{\omega_{+}^{2}}
 \right)}.
   \label{a}
\end{equation}
Finally, eq.\ (\ref{omega}) yields $1/\tilde{r}$ according to the
relation (\ref{1/r}). The unknowns $\tilde{M}$,
$\tilde{a}/\tilde{M}$ and $\tilde{r}/\tilde{M}$ are reached by
obvious combinations.

The results of this approximation are illustrated in Table~1 for several
typical $a$ and $r$, together with the respective values of the
observable quantities. Here, $a$ and $r$ are assumed to be given (in
units of $M$; see the text above for conversions to physical units). For
a prograde orbit of the secondary, $\omega_+$ and
$\Omega_{\infty}(\omega_+)$ are the corresponding azimuthal and
latitudinal angular frequencies, $\delta=\delta(\omega_+)$ is the
dimensionless parameter defined by eq.~(\ref{delta}). From these
quantities one computes the estimates $\tilde{a}$ and $\tilde{r}$ (in
units of $\tilde{M}$) using our approximate eqs.\ (\ref{r})--(\ref{a}).
Notice that acceptable results are reached for $r\gta10M$ where
approximative values are close to the simulated exact ones. Accuracy
decreases if the secondary orbits too close. On the other hand, too
distant orbits are also unsuitable: dragging effects weaken quickly with
distance from the centre and the values of $\omega_+$ and
$\Omega_{\infty}$ become too close to each other. In other words, the
minimum sampling rate (inverse of the Nyquist critical frequency; Press
et al.\ 1992) is restricted by the difference
$\tau=\Omega_{\infty}^{-1}-\omega_+^{-1}$, requiring that both
frequencies can be safely distinguished in the observed signal. For
example, for $M=10^8M_\odot$, maximum rotation ($a=1$), and $r=20$
gravitational radii, one can find (from Table~1) $\tau=1.8$, which in
Fig.~1 corresponds to time interval $T-t=2\pi{M}c^{-1}\tau=1.5$~hr.

\section*{References}
\small
\parskip 2pt
\re
   Adams, F.~C.,  Watkins, R.\ 1995, ApJ 451,~314
\re
   Blandford, R.~D.,  McKee, C.~F.\ 1982, ApJ 255,~419
\re
   Bracco, A., Provenzale, A., Spiegel, E.~A.,  Yecko, P. 1998,
   in Theory of Black Hole Accretion Disks, eds.\ M.~A. Abramowicz,
   G.~Bj\"ornson, J.~E. Pringle
   (Cambridge University Press, Cambridge) 
\re
   Bromley, B.~C., Chen, K., Miller, W.~A.\ 1997, ApJ 475,~57
\re
   Charles, P.~A.\ 1997,
   in Proc.\ 18th Texas Conference (World Scientific, Singapore), in press
\re
   de Felice, F.,  Usseglio-Tomasset, S.\ 1996, Gen.\ Rel.\ Grav. 28,~179
\re
   Evans, C.~R., Kochanek, C.~S.\ 1989, ApJ 346,~L13
\re
   Fanton, C., Calvani, M., de Felice, F., \v{C}ade\v{z},~A.
   1997, PASJ 49,~159
\re
   Ivanov, P.~B., Igumenshchev I.~V.,  Novikov I.~D. 1998, ApJ 507,~131
\re
   Iwasawa,~K., Fabian, A.~C., Brandt, W.~N., Kunieda,~H.,
   Misaki,~K., Reynolds, C.~S., Terashima,~Y. 1998,
   MNRAS 295,~L20
\re
   Karas, V. 1997, MNRAS 288,~12
\re
   Karas, V.,  Vokrouhlick\'y, D.\ 1994, ApJ 422,~218
\re
   Kato, S. 1990, PASJ 42,~99
\re
   Kato, S., Fukue, J.,  Mineshige, S. 1998, Black-hole Accretion
   Disks (Kyoto University Press, Kyoto)
\re
   Kormendy, J.,  Richstone, D.\ 1995,
   ARAA 33,~581
\re
   Liu, F.~K., Liu, B.~F., Xie, G.~Z.\ 1997, AAS 123,~569
\re
   Luminet, J.-P., Marck, J.-A.\ 1985, MNRAS 212,~57
\re
   Martocchia, A., Karas, V., Matt, G.\ 1999, MNRAS, submitted
\re
   Misner, C.~W., Thorne, K.~S., Wheeler, J.~A.\ 1973,
   Gravitation (Freeman, San Francisco)
\re
   Nandra, K., George, I.~M., Mushotzky, R.~F., Turner, T.~J.,
   Yaqoob, T.\ 1997, ApJ 476, 70; ApJ 477,~602
\re
   Narayan, R., Garcia, M.~R., McClintock, J.~E.\ 1997, ApJ 478,~L79
\re
   Nowak M.~A., Wagoner R.~V.\ 1992, ApJ 393,~697
\re
   Okazaki, A.~T., Kato, S., Fukue,~J.\ 1987, PASJ 42,~99
\re
   Press, W.~H., Teukolsky, S.~A., Vetterling, W.~T., Flannery, B.~P.\
   1992, Numerical Recipes (Cambridge University Press, Cambridge)
\re
   Rees, M.~J.\ 1988, Nature 333,~523
\re
   Rees, M.~J.\ 1998, in Black Holes and Relativistic Stars,
   ed.\ R.~M. Wald (Univ.\ Chicago Press, Chicago)
\re
   Semer\'ak, O., de Felice, F.\ 1997, Class.\ Quantum Grav.\ 14, 2381
\re
   Stella, L.\ 1990, Nature 344,~747
\re
   Sundelius,~B., Wahde,~M., Lehto, H.~J.,  Valtonen, M.~J. 1997,
   ApJ 484,~180
\re
   Syer, D., Clarke, C.~J.,  Rees, M.~J.\ 1991, MNRAS 250,~505
\re
   Young A.~J., Ross R.~R., Fabian A.~C.\ 1998, MNRAS 300,~L11
\re
   van der Klis, M.\ 1997, in Proc.\ NATO ASI, The Many Faces of
   Neutron Stars;
   astro-ph/9710016
\re
   Villata,~M., Raiteri, C.~M., Sillanp\"a\"a,~A., Takalo, L.~O.
   1998, MNRAS 293,~L13
\re
   Vokrouhlick\'y, D.,  Karas, V.\ 1993, MNRAS 265,~365
\re
   Vokrouhlick\'y, D.,  Karas,~V.\ 1998, MNRAS 298,~53
\re
   Wilkins, D.~C.\ 1972, Phys.\ Rev.\ D 5,~814
\re
   $\dot{\rm{Z}}$ycki P.~T., Done C., Smith D.~A.\ 1998, ApJ 496,~L25


\protect\newpage
\onecolumn




\slugcomment{To appear in The Astrophysical Journal}
\lefthead{V.~Karas}
\righthead{Twin peaks separation in sources with kHz QPO\ldots}

\hyphenation{Schwarz-schild}
\hyphenation{Szusz-kie-wicz}

\title{Twin peaks separation in sources with kHz QPOs
 caused by orbital motion\footnote{To appear in The Astrophysical Journal 
 (1999)}}
\author{Vladim\'{\i}r Karas\altaffilmark{1}}
\altaffiltext{1}{Astronomical Institute,
                 Charles University Prague,
                 V~Hole\v{s}ovi\v{c}k\'ach~2,
                 CZ-180\,00 Praha~8,
                 Czech Republic;
                 vladimir.karas@mff.cuni.cz}

\vspace*{3cm}

The expected range of kilohertz quasi-periodic oscillations (kHz QPOs)
is studied regarding low-mass X-ray binaries with twin peaks in
their Fourier spectra. The twin peaks are interpreted as combination of
epicyclic and precesion frequencies in Schwarzschild metric. If the
pairs are caused by general relativistic orbital motion of gaseous
clumps around a neutron star, then moderate size of the clumps and
eccentricity of their orbits can account for the non-constant peaks
separation reported in Scorpius X-1 and 4U 1608-52. The neutron stars
mass comes out in the range 1.6--1.9 solar masses. Variation in the
orbital parameters of the clumps translates into an expected range in
the plot of pair separation versus frequency, typically showing a
decreasing slope.


\noindent
Keywords: {X-rays: stars --- stars: individual (Sco X-1, 4U 1608-52)
--- relativity}

\twocolumn
\setcounter{equation}{0}
\section{Introduction}
Accreting low-magnetic field neutron stars exhibit complex
behaviour including X-ray variability on millisecond time-scales;
see van der Klis (1998) for a review of millisecond phenomena in
low-mass X-ray binaries, and its recent update (van der Klis 1999).
Cf.\ M\'endez (1999) for a recent summary of {\em{RXTE}} observational
results and for further references. Here we concentrate on the widely
discussed twin kHz peaks which so far have been reported in the Fourier
power spectra of nearly twenty objects. The importance of this feature
follows from the fact that the positions of the two peaks (and of the
third, burst oscillation frequency), their large coherence and variation
with the count rate restrict viable models of the accreting body and of
the geometry of the system. Also, it turns out that effects of general
relativity cannot be ignored (Klu\'zniak 1998; Stella \& Vietri 1998;
Zhang et al.\ 1998).

The approximately constant separation of the twin QPO peaks indicates
that some kind of beat-frequency relation (Miller, Lamb, \& Psaltis
1998) can explain these pairs, although the simplest models have been
challenged by the discovery of QPO sources in which the separation of
the peaks decreases as their frequency increases: Sco X-1 (van der Klis
et al.\ 1997) and 4U 1608-52 (M\'endez et al.\ 1998). Other sources are
consistent with this trend too, e.g., 4U 1735-44 (Ford et al.\ 1998) and
4U 1728-34 (M\'endez \& van der Klis 1999). An inspiring alternative
has recently been proposed by Stella \& Vietri (1999) who introduce
frequencies of the orbital motion around a compact body, namely orbital
frequency, epicyclic frequency, and their difference. Here, this idea
will be further studied. In particular, the range of expected frequency
variation of the twin peaks will be estimated assuming that the feature
arises from the motion of clumps with small but finite size, orbiting
along eccentric orbits. The aim is to examine the main difference of
this interpretation against simple constant-separation scenario by
quantifying the expected separation as a function of peaks frequency.
The mass of the neutron star stands as a parameter of the procedure in
which predicted frequencies are fitted to data. It will be shown that
the picture is indeed in agreement with the decreasing difference of the
QPO peaks in Sco X-1 and 4U 1608-52. Even more importantly, the model is
testable in principle, and it can be rejected when predicted frequencies
of the peaks are found inconsistent with more accurate data in future.

\section{Frequencies relevant for the twin peaks}
Gravitational field is described in terms of Schwarzschild metric
(specified by the mass $M$ of the body which is taken to be of the order
of two solar masses). We will first write down expressions for the two
fundamental frequencies driving the orbital motion of blobs, which are
assumed to follow free trajectories (Chandrasekhar 1983). This is the
simplest situation which of course will have to be abandoned in a more
realistic model when the influence of radiation and of magnetic forces is
taken into account, or when the gravitational field itself contains higher
multipoles. Trajectories in a spherical gravitational field of
Schwarzschild metric are planar but, in general, they do not form a
closed curve in space. In the case of bound stable orbits satisfying
the condition
\begin{equation}
 \label{cond}
 1-6\mu-2{\mu}e\geq0, \quad \mu{\equiv}Ma^{-1}\left(1-e^2\right)^{-1},
\end{equation}
one can associate eccentricity $e$ and semimajor axis $a$ with
corresponding values of pericentre ($r_p$) and apocentre ($r_a$)
distances, $e=(r_a-r_p)/(r_a+r_p),$ $a=r_p/(1-e)$. The orbit
can be characterized by the pericenter shift $\delta\phi_p$
during period $P$ of time which elapses between subsequent passages
through pericentre of the orbit. (Here, temporal quantities are
expressed with respect to a distant observer at rest and measured in
geometrized units, while azimuthal angles are defined in the orbital
plane in the standard manner). Both $\delta\phi_p$ and $P$ can be written
in the form of integrals over relativistic true anomaly ${\bar{\chi}}$,
\begin{equation}
 \label{deltaphi}
 \delta\phi_p=\int_{0}^{2\pi}
 \left[1-2\mu\left(3+\eta\right)\right]^{-1/2}{\rm{d}}{\bar{\chi}} - 2\pi,
\end{equation}
\begin{eqnarray}
 \label{P}
 P & = & {\cal{M}} \int_{0}^{2\pi}
 \left(1+\eta\right)^{-2}
 \left[1-2\mu\left(3+\eta\right)\right]^{-1/2}
 \nonumber \\
  & & \times
  \left[1-2\mu\left(1+\eta\right)\right]^{-1}{\rm{d}}{\bar{\chi}},
\end{eqnarray}
where $\eta=e\cos{\bar{\chi}}$,
${\cal{M}}=M\mu^{-3/2}\sqrt{(2\mu-1)-(2{\mu}e)^2}$. The frequency of
radial oscillations (epicyclic frequency) is thus $\nu_r=P^{-1}$, and
precession frequency $\nu_p=\delta\phi_p/(2{\pi}P)$. Combination of
these two frequencies gives the corresponding azimuthal frequency,
$\nu_\phi-\nu_p=\nu_r$.\footnote{\protect\small{}These quantities are 
introduced in
standard manner, similar to Stella \& Vietri (1999). Notice, however,
that our definition of $\nu_p$ differs slightly from their corresponding
relation for $\nu_{\it{per}}$, because here it is expressed in terms of
the integral over the {\em{whole range\/}} of $2\pi$ in true anomaly.
This is a proper and natural definition although the difference is only
moderate for low-eccentricity orbits, considered in these papers. For
eccentric orbits around $6M$, azimuthal and precession frequencies are
comparable while frequency of radial oscillations is much less.} Eqs.\
(\ref{deltaphi})--(\ref{P}) can be written in terms of elliptic
integrals (Karas \& Vokrouhlick\'y 1994). Frequencies are then given in
a more elegant form, although direct numerical evaluation of integrals
(\ref{deltaphi})--(\ref{P}) is equivalent for practical purposes.

Several instability mechanisms have been identified which may set clumps
of gaseous material onto an eccentric orbit down from the inner edge of
an accretion disk or from the sonic point. Miller et al.\ (1998) remark
that there may even be hundreds of clumps accumulated near the sonic
point which can subsequently be set on an inspiralling trajectory. There
is an interesting possibility which concerns freely moving clumps: a
{\em{steady eccentric ring}\/} may be formed from individual blobs, surviving a
number of orbital periods and rotating with precession frequency $\nu_p$
(Bao, Hadrava, \& {\O}stgaard 1994). This presumes that clumps of matter
are set on the eccentric orbit ($a$, $e$ given) near the apocentre, one
after another. Radiation from illuminated or self-radiating clumps is
modulated at orbital and precession frequencies, and their differences
also appear in the harmonic content of the observed signal. We thus
examined the possibility of interpreting the difference between the two
peaks as $\Delta\nu=\nu_r$, while the upper twin peak can be interpreted
either as $\nu=\nu_\phi$ (as in Stella \& Vietri 1999), or alternatively
as $\nu=\nu_p$. As we show later on, both interpretations lead to
acceptable agreement with data on the peak frequencies in Sco X-1 and 4U
1608-52, but the latter one corresponds to the masses of the neutron
star which are lower by about $0.3M_{\odot}$.

\begin{figure}[t]
\centering\leavevmode
\epsfxsize=\hsize
\epsfbox{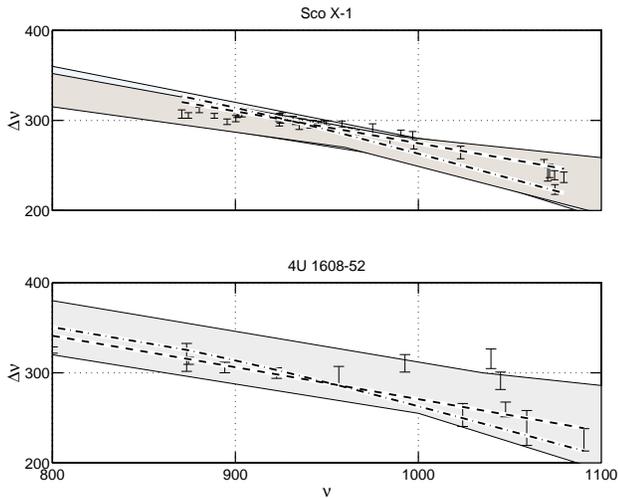}
\caption{\protect\small{}
The best fits in the plot of the pair separation
$\Delta\nu(\nu)$ for $\nu=\nu_\phi$ (dot-dashed) and $\nu=\nu_p$
(dashed), respectively. For the latter case, shaded regions indicate the
expected variation $\Delta\nu$ as function of the upper peak frequency
$\nu$. Notice that interpretation of peaks in terms of epicyclic and
precession frequencies with reasonable variation of free parameters
covers the data and captures the trend of decreasing $\Delta\nu(\nu)$.
Frequencies are in Hertz. See the text for further details.
\label{fg1}}
\end{figure}

\begin{figure}[t]
\centering\leavevmode
\epsfxsize=\hsize
\epsfbox{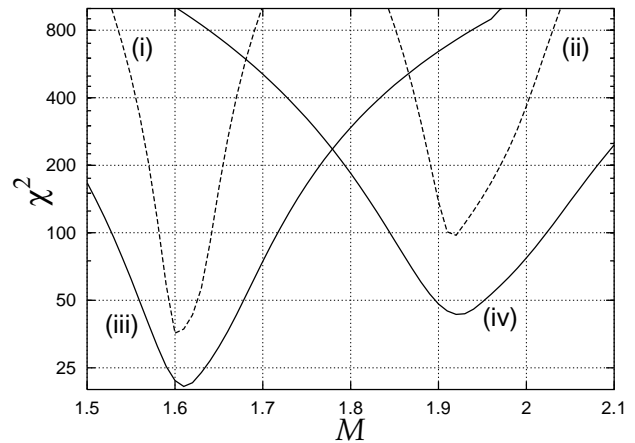}
\caption{\protect\small{}
Standard $\chi^2$ test of the best-fit mass parameter
$M$ (units of solar mass). In this figure, curves (i) and (ii) are for Sco
X-1; (iii) and (iv) are for 4U 1608-52. Higher values of the mass, (ii)
and (iv) ($M=1.92M_{\odot}$), correspond to $\nu=\nu_\phi$, while lower
values, (i) and (iii) ($M=1.61M_{\odot})$, correspond to $\nu=\nu_p$ for
both sources. \label{fg2}}
\end{figure}

The frequencies were evaluated for orbits satisfying condition
(\ref{cond}). (Notice that pericenter of eccentric orbits can be below
the marginally stable orbit of circular motion, which is at $r=6M$ in
Schwarzschild metric.) There are three free parameters of the model, $M$, $a$,
and $e$, which can be varied to determine the best fit to the
observational relation $\Delta\nu(\nu)$. Given these parameters, we
still allow for a small variation of the integration range in
${\bar{\chi}}$, Eqs.\ (\ref{deltaphi})--(\ref{P}), which represents
uncertainty in the clumps size $d$ of the order of 3\% of $a$. In
other words, the bounds of the interval $\langle0,2\pi\rangle$ for true
anomaly were varied by $\delta{\bar{\chi}}(d)$ corresponding to
finite-size of clumps. For this purpose one only needs to know the
relation between a small variation in true anomaly and the corresponding
variation in azimuthal angle:
$\delta{\bar{\chi}}=
\xi^{1/2}\left[1-k^2\cos^2({\bar{\chi}}/2)\right]^{1/2}\delta\phi$,
where $k^2=4{\mu}e\xi^{-1}$ and $\xi=1-6\mu+2{\mu}e$. Near apocentre,
${\bar{\chi}}\approx\pi$, the shape of the trajectory is very close to
its osculating ellipse in this point.

Figure \ref{fg1} shows the range which can be covered by this model in
plots of $\Delta\nu(\nu)$ for Sco X-1 and 4U 1608-52. These sources are
known to exhibit decreasing separation $\Delta\nu$ when $\nu$ increases
(data taken from van der Klis et al.\ 1997, and M\'endez et al.\ 1998).
The best fit is plotted for the two interpretations of the upper-peak
frequency: (i)~$\nu=\nu_\phi$ (dot-dashed line) and (ii)~$\nu=\nu_p$
(dashed), respectively. We used standard $\chi^2$ statistics to
determine the best-fit values of the parameters. The case (i)
corresponds to Stella \& Vietri (1999), and it yields the neutron star
mass equal to $1.92M_{\odot}$, incidentally for both sources, in good
agreement with their fit (apart from small difference in calculation of
relevant frequencies, which translates to a 2\% decrease of the best-fit
mass; cf.\ footnote 2). $Q$ probability is $10^{-5}$, defined in terms
of $\Gamma$ function and evaluated with deviations as in Fig.\
\ref{fg1} (Press et al.\ 1992). The fit can be slightly improved by
varying $e$, as suggested by Stella \& Vietri, and one could also use a
robust fit. The case (ii) results in a graph of $\Delta\nu(\nu)$ which
also decreases monotonically, but the slope is flatter. We will
concentrate on this case since the former one has already been examined.
The best-fit mass is now significantly less, $1.61M_{\odot}$ ($Q=0.4$).
The dark shaded region in Fig.~\ref{fg1}a corresponds to variation in
$\delta{\bar{\chi}}$ around the best fit. The shading covers a part of
the $(\Delta\nu,\nu_p)$ plane where the best fit is moved by such
variation. Although all parameters are interconnected in the equations
for frequencies, one can say that $M$ mainly determines the absolute
value of upper-peak frequency $\nu$, while $a$ and $e$ affect the
sloping stripe of $\Delta\nu(\nu)$. Typically, in our calculations $a$
ranges between $6M$--$10M$, while $r_p$ can plunge down to $4M$,
determining the corresponding eccentricity $e\lesssim0.4$. Keeping $a$
fixed, the slope increases as $e$ decreases. An almost identical area of
the graph can be covered by forcing the mass to a slightly higher value,
$M=1.70M_\odot$ (indicated by somewhat lighter shading in Fig.\ \ref{fg1}
for Sco X-1). In the latter case, however, $\delta{\bar{\chi}}$ has to be 
varied in a range corresponding to somewhat larger size of the clumps, from 
3 to 6 per cent of $a$. For 4U 1608-52, the larger scatter in the data
translates to a requirement on clumps size up to $\approx8$\% of $a$.

Let us remark that non-zero size of the spots is inevitable if they are
to modulate visibly the observed radiation from the source for long
periods of time. On the other hand, shearing motion destroys large spots
and sets an upper limit to their sizes. {\em{If}\/} individual elements
of the spots followed free Keplerian trajectories on their neighbouring
orbits, the above-mentioned 3\% diameter (in units of $a$) would
determine maximum spot size, which is still consistent with small
observed widths of the peaks.\footnote{\protect\small{}
We thank the referee for this
point. A particular example of a diminishing signal from spots which decay
by shearing motion in the course of several revolutions was shown by
Karas, Vokrouhlick\'y, \& Polnarev (1992).} Indeed, the peaks' width
increases with diameter of the sheared spots, and it should stay in
agreement with the value of $d$, which has been determined here from
scatter of data around the best fit in Fig.\ \ref{fg1}. Notice,
however, that several physical mechanisms have been proposed where the
spots in accretion disks could survive many orbital periods, in spite of
their finite sizes. In this respect, formation of vortices was suggested
by Abramowicz et al.\ (1992), and examined by Adams \& Watkins (1995) and
Bracco et al.\ (1999). It is speculated that these vortices durate over
sufficient number of local orbital periods of the vortex center, giving
rise to distinct features in observed power spectrum of the source.
However, the origin of vortices is an open question so far. We thus
remain on a phenomenological level of bright spots.

Figure \ref{fg2} shows graphs of $\chi^2$, which were obtained by
fitting $a$ and $e$, while $M$ was kept a fixed value within the range
1.5--2.1 solar masses and $\delta\bar{\chi}=0$. Position of the minimum
of $\chi^2(M)$ suggests the most probable mass. There are two different
values of $M$ clearly distinguished from each other, which correspond to
the two cases of fitting described above, (i)~$M\approx1.9M_{\odot}$ and
(ii)~$M\approx1.6M_{\odot}$, although one must be cautious by
interpreting such a graph when accuracy of data and the number of data
points are limited. The fit can be further improved by factor $\approx2$
by varying $\delta{\bar{\chi}}$ as described above.

Other physical processes, which have been neglected in the present text,
introduce corrections to the expected range of frequencies, even if the
main interpretation of relativistic orbital motion holds. Namely, one
wants to estimate corrections which are caused by difference in the
gravitational field of a rotating body in comparison with non-rotating
Schwarzschild metric adopted here. It follows from the discussion of
Morsink \& Stella (1999) that no realistic equation of state for the
neutron star can account for a change in kHz QPO frequencies more than
10\%. An analogous conclusion holds for motion in Kerr metric with
maximum rotation. It is, however, important to note that inclined orbits
around a rotating body are not planar. This is due to Lense-Thirring
precession which may give rise to another fundamental frequency from a
tilted accretion disk (Stella \& Vietri 1998; Cui, Zhang, \& Chen 1998).
We refer to Merloni et al.\ (1999) for detailed evaluation of these
fundamental frequencies of spherical orbits, and for corrections to Cui
et al.\ (1998) paper. Even in the case of non-spherical perturbations of
the orbits, which should occur in realistic situations, one can
introduce a {\em{mean pericenter shift as an average}\/} taken over many
orbits, or, equivalently, over a single orbit of many individual clumps
described by the probability distribution of their orbital parameters.
See Karas \& Vokrouhlick\'y (1994) for such discussion of fundamental
frequencies of eccentric orbits in Kerr metric. Possible appearence of
averaged frequencies in the power spectrum is illustrated there too.
Notice that small non-zero eccentricity of the orbits of the clumps
gives rise to variations of the expected peaks' frequencies about the
mean value.

\section{Conclusion}
We have verified that observed frequencies of the twin kHz QPO peaks can
be reproduced by assuming modulation of the observed signal by the two
fundamental frequencies of orbital motion. An alternative estimate of
the neutron star mass in Sco X-1 is consistent with
$M\approx(1.6$--$1.7)M_\odot$, somewhat less than $\approx1.9M_\odot$
derived by Stella \& Vietri (1999). The difference arises from
interpretation of the upper peak ($\nu_p$ vs.\ $\nu_{\phi}$), and both
possibilities still remain consistent with observational data. For
orbits with negligible eccentricity we obtained identical results as
those illustrated in their Fig.~1, but the present scheme obviously
requires non-zero eccentricity to show up the precession frequency in
the signal. In both cases the expected range of the pair's frequency
corresponds to orbital motion around a $\lesssim2M_\odot$ compact
object, and their separation decreases with increasing frequency. The
decrease is faster for less eccentric orbits (within the relevant range
of frequencies, i.e.\ $700\,{\rm{Hz}}\lesssim\nu\lesssim1200\,{\rm{Hz}}$
for the upper peak). An appealing feature in the present highly
simplified treatment is that the expected range, as indicated by the
shading in Fig.~\ref{fg1}, is very sensitive to the value of the
central mass, and it moves away from the scale of our graphs when $M$ is
changed by $\approx0.3M_\odot$, independently of other free parameters.
Although the orbital parameters $e$ and $\delta{\bar{\chi}}$ are not
strictly determined in the present treatment, it is worth mentioning
that one cannot fiddle their values to obtain an increasing slope of
$\Delta\nu(\nu)$; this might be possible only if $\nu\lesssim700\,$Hz
and/or $M\lesssim1.5M_\odot$. This argument can be used to
{\em{exclude\/}} the present interpretation in other sources. For
example, in the case of 4U 1728-34 (M\'endez \& van der Klis 1999) the
available data do not allow a consistent fit by the above described
procedure, although the peaks separation is again decreasing function of
frequency.

\acknowledgments
The author acknowledges referee's suggestions regarding finite size of
the blobs and critical comments on the first version of this paper.
Hospitality of the International School for Advanced Studies in Trieste
and support from the grants GACR 205/\lb{2}97/\lb{2}1165 and
202/\lb{2}99/\lb{2}0261 is also acknowledged.


\section*{References}
\parskip 0pt
\re
   Abramowicz, M.~A., Lanza,~A., Spiegel, E.~A., Szuszkiewicz,~E.
   1992, Nature, 356,~41
\re
   Adams, F.~C., Watkins, R.\ 1995, \apj, 451, 314
\re
   Bao, G., Hadrava,~P., {\O}stgaard,~E.\ 1994, \apj, 425,~63
\re
   Bracco, A., Provenzale,~A., Spiegel, E.~A., Yecko,~P.\ 1999,
   in Theory of Black Hole Accretion Disks, eds.\ M.~A.\ Abramowicz,
   G.~Bj\"ornson, J.~E.\ Pringle (Cambridge University Press: Cambridge)
\re
   Chandrasekhar, S.\ 1983, The Mathematical Theory of Black Holes
   (Clarendon Press: Oxford)
\re
   Cui, W., Zhang, S.~N., Chen,~W. 1998, \apjl, 492,~L53
\re
   Ford, E.~C., van der Klis, M., van Paradijs, J., M\'endez, M.
   1998, \apjl, 508,~L155
\re
   Karas, V., Vokrouhlick\'y,~D.\ 1994, \apj, 422,~218
\re
   Karas, V., Vokrouhlick\'y,~D., Polnarev,~A.\ 1992, \mnras, 259,~569
\re
   Klu\'zniak, W.\ 1998, \apjl, 509,~L37
\re
   M\'endez, M.\ 1999, in the electronic Proceedings of the 19th Texas
   Symposium (Paris 1998), to appear
\re
   M\'endez, M., van der Klis,~M.\ 1999, \apjl, in press
\re
   M\'endez, M., van der Klis,~M., Wijnands,~R., Ford, E.~C.,
   van Paradijs,~J., Vaughan, B.~A.\ 1998, \apjl, 505,~L23
\re
   Merloni, A., Vietri, M., Stella, L., Bini, D.\ 1999, \mnras, 304, 155
\re
   Miller, M.~C., Lamb, F.~K., Psaltis,~D.\ 1998, \apj, 508,~791
\re
   Morsink, S., Stella,~L. 1999, \apj, in press
\re
   Press W.~H., Teukolsky S.~A., Vetterling W.~T., Flannery B.~P.\ 1992,
   Numerical Recipes (Cambridge University Press: Cambridge)
\re
   Stella, L., Vietri, M.\ 1998, \apjl, 492,~L59
\re
   Stella, L., Vietri, M.\ 1999, \prl, 82,~17
\re
   van der Klis, M.\ 1998, in Proceedings of the NATO ASI Ser.\ C, vol.\
   515, The Many Faces of Neuton Stars, p.~337
\re
   van der Klis, M.\ 1999, in Proceedings of the Third
   William Fairbank Meeting on the Lense-Thirring Effect, in press
\re
   van der Klis, M., Wijnands, R.~A.~D., Horne,~K., Chen,~W.\ 1997,
   \apjl, 481,~L97
\re
   Zhang, W., Smale, A.~P., Strohmayer, T.~E., Swank, J.~H.\ 1998,
   \apjl, 500,~L171


\protect\newpage
\onecolumn



\markboth{V.\ Karas}{Power Spectrum of Orbiting Clumps}


\centerline{\Large\bf Quasi-Periodic Features due to Clumps Orbiting
around a Black Hole\footnote{To appear in Publ.\ Astron.\ Soc.\ Japan 
(1999)}}

\vspace*{3cm}

{Vladim\'{\i}r {\sc{}Karas}
\\
{\it{}Astronomical Institute, Charles University Prague,
V~Hole\v{s}ovi\v{c}k\'ach~2, CZ-180\,00~Praha, Czech Republic}
\\
{\it{}E-mail: vladimir.karas@mff.cuni.cz}}

\vspace*{2cm}

{The expected Fourier power-spectrum density is examined with
regard to the shape and quality factor $Q$ of features which are
predicted to appear in a signal from clumps orbiting near a marginally
stable orbit of a compact body. It is shown that clumps distributed
randomly in a narrow range of radii produce quasi-periodic oscillations
with $Q\approx10^2$, similar to those observed in many low-mass X-ray
binaries.}

\noindent
Keywords: {X-rays: sources --- Accretion disks --- Black holes}


\twocolumn
\setcounter{equation}{0}
\section{Introduction}
Different types of astronomical objects exhibit periodic modulation of
their signal, which contains information about the physical properties
of the source. Accreting low-mass X-ray binaries (LMXRBs) with
quasi-periodic oscillations (QPOs) represent a particular example which
has attracted much interest (for recent review see van der Klis 1999).
While the origin of the QPOs is not fully understood, virtually all
viable models introduce inhomogeneities (clumps of gas) in an accretion
disk orbiting near to a compact neutron star or a black hole. The
features which are clearly visible in the Fourier power spectra can be
attributed to Doppler and lensing effects modulating the observed signal
which comes from those rapidly moving clumps. Oscillations also arise
due to some kind of beat frequency with neutron star spin (Miller et
al.\ 1998). The QPO frequencies can be rather high, up to
$\approx1300\,$Hz, which is comparable with dynamical time-scale at the
marginally stable orbit of a stellar-mass object, $R_{\rm{ms}}$,
suggesting distances of only several gravitational radii from the
central body. Similar oscillations have been reported in X-rays from
some active galactic nuclei (AGN; Papadakis, Lawrence 1995). Here, the
central masses are higher by six orders of magnitude and more, so that
dynamical time is a few hours. Observational data are somewhat scarce in
the case of AGN, compared to Galactic sources, because the time-scales
are longer and the corresponding features in the power spectrum are less
sharp than in the case of LMXRBs. Still, it appears that orbiting
inhomogeneities contribute to the observed X-ray variability of AGN
although their presence does not explain the entire variability spectrum
on a purely kinematic basis. The case of AGN is an issue of much debate,
and other (magnetohydrodynamic) processes are clearly important
(Abramowicz et al.\ 1991; Karas 1997).

We examined the expected variability power spectrum from the
distribution of clumps orbiting around a compact body. In this note we
employ a phenomenological description of the clumps, which has been
adopted, in some form, by present models of sources with QPOs. These
inhomogeneities are visible in a narrow zone of a few gravitational
radii (where luminosity of the accretion disk is maximum) with the
lifetime greatly exceeding the corresponding orbital periods. The
above-mentioned assumption concerning the strictly localized radial
distribution of the clumps is one of the distinctions of the present
work from analogous studies in which the short-term featureless X-ray
variability of AGN has been explored in terms of a clumpy accretion
disk. Bao \& {\O}stgaard (1995) assumed a broad range of radii where the
clumps exist in the disk and modulate observed signal (of the order of
$10^2$ gravitational radii), resulting in a power-law form of the
Fourier power-spectrum density; in these circumstances individual peaks
are smeared away. The fundamental frequency of orbital motion would
stand out if all of the clumps were aligned along a ring with a strictly
determined radius. Because this will not be the case in more realistic
models, it appears relevant to ask how the width of the feature in the
power spectrum is modified when the clumps are spread over a small, but
non-zero, range of radius. There are at least two reasons why a
localized distribution of clumps should be taken into account. First,
several instability mechanisms have been proposed which could operate
within distances of several $R_{\rm{g}}$ from the center of relativistic
accretion discs and result in trapped perturbations there (Kato, Fukue
1980; Markovi\'c, Lamb 1998). Second, a non-axisymmetric structure of
the disk surface could be best visible from the innermost parts of the
disk where its radiation and irradiation are much stronger than farther
from the center (e.g., Matt et al.\ 1991; Pariev, Bromley 1998). Peaks
in the Fourier spectrum corresponding to the localized distribution of
the clumps can be characterized by the quality factor $Q=\nu/\Delta\nu$
(peak's centroid frequency divided by its full width at half maximum),
and by harmonic content of the signal.

We considered different distributions of the clumps, as characterized by
their number density and local emissivity. Here we report the main
conclusion of this study, namely, we show that maximum expected values
of parameter $Q$ are of the order of one hundred and they can be as
large as $3\times10^2$. We will also show typical examples of
the expected power spectrum with its harmonic content which is rather
strong at medium and large (i.e.\ edge-on) inclination.

\begin{fv}{1}
{0pc
\epsfxsize=0.97\hsize\epsfbox{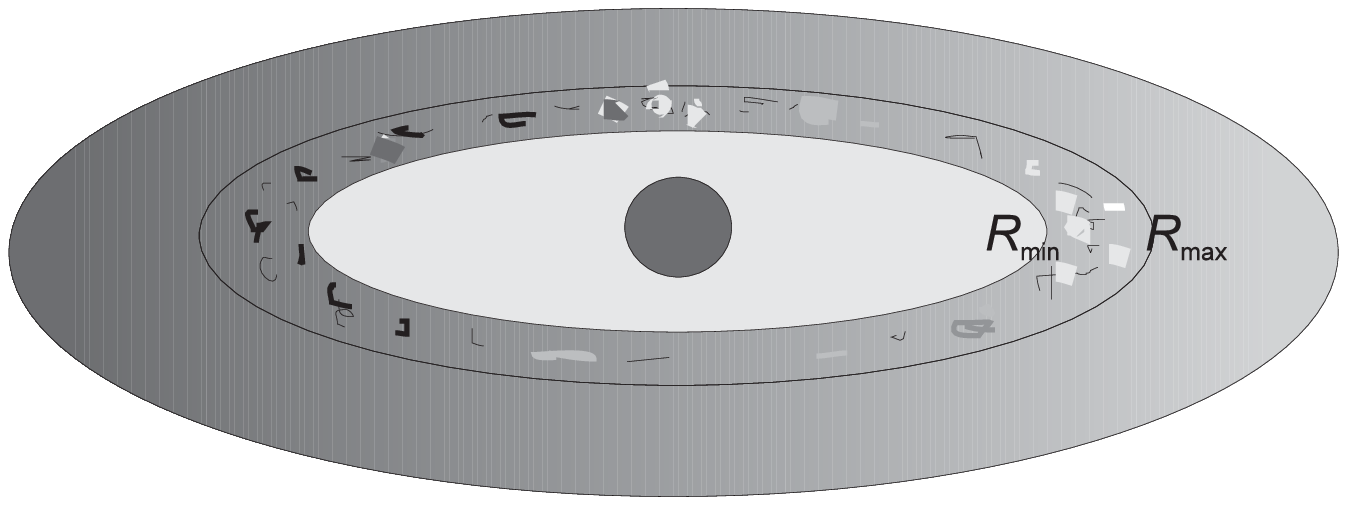}
}
{\protect\small{}In 
this model, clumps are distributed in a zone of several
gravitational radii with their observed luminosity influenced by Doppler
and lensing effects.
\label{fig1}}
\end{fv}

\begin{fv}{2}
{0pc
\epsfxsize=0.93\hsize\hspace*{0.6cm}\epsfbox{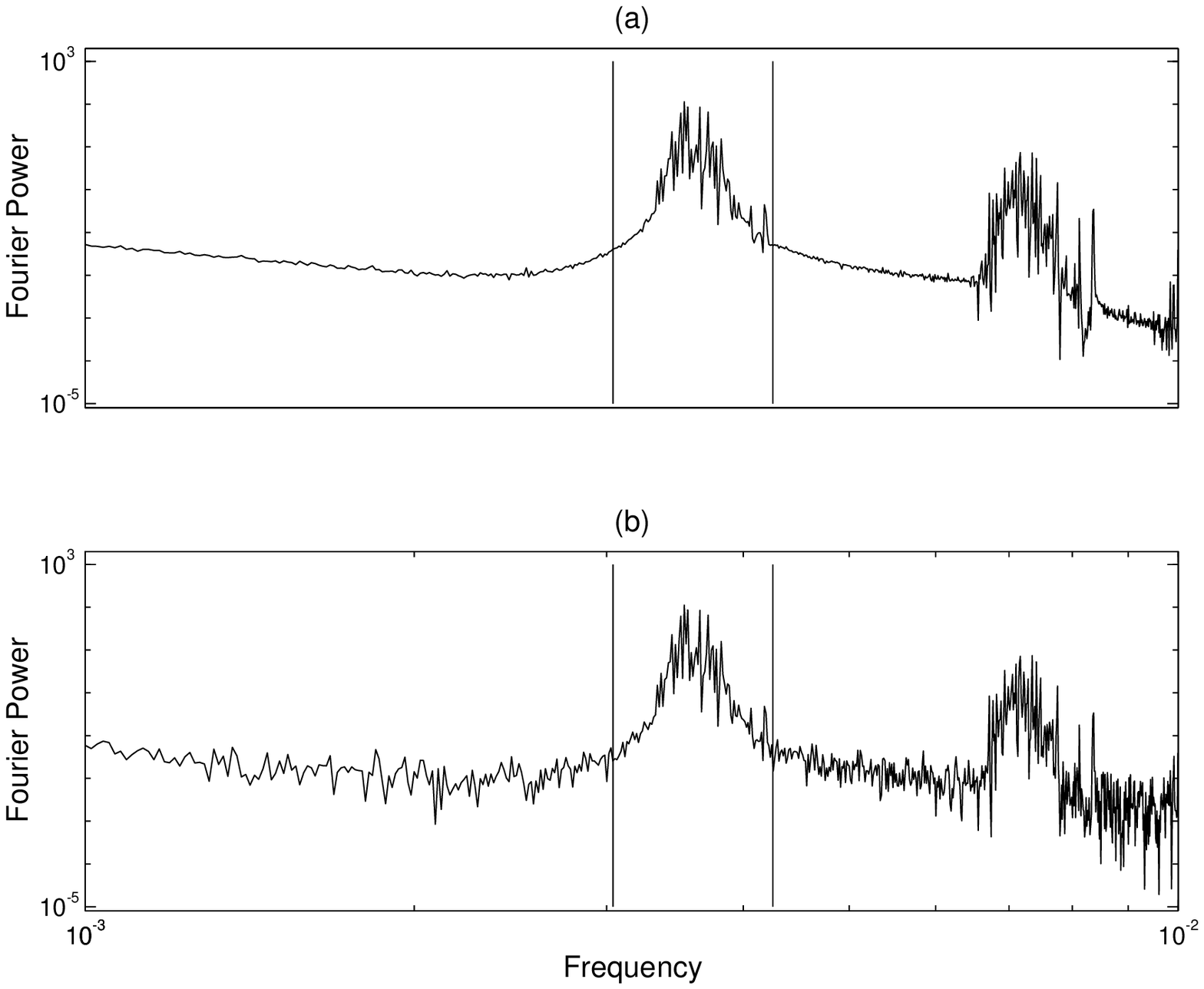}
}
{\protect\small{}Logarithmic 
plot of the expected power-spectrum density. The two panels
correspond to: (a)~pure superposition of light curves of mutually
independent sources orbiting just above the marginally stable orbit;
(b)~an additional contribution of the high-frequency noise. Vertical
bars indicate Keplerian frequencies at the edges of the clumps radial
distribution, i.e.\ the range of fundamental frequencies
$\langle\nu_{\rm{k}}(R_{\rm{min}}),
\nu_{\rm{k}}(R_{\rm{max}})\rangle$. Here, the frequency 
(horizontal axis) is in geometrized units, while the
power spectrum (vertical axis) has an arbitrary scale. See the text for
details.
\label{fig2}}
\end{fv}

\begin{fv}{3}
{0pc
\epsfxsize=0.93\hsize\hspace*{0.6cm}\epsfbox{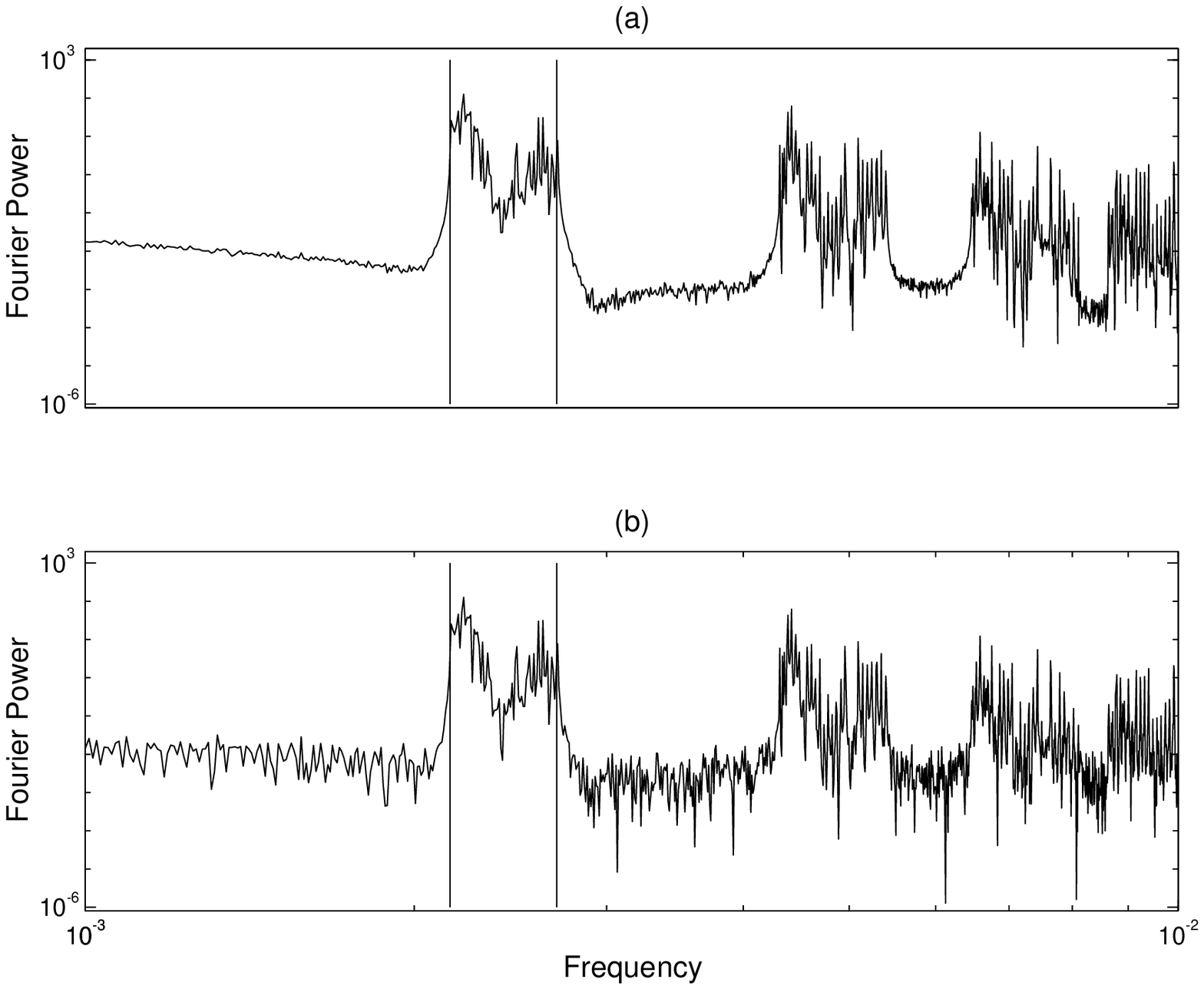}
}
{\protect\small{}As 
in figure 2, but for the radial distribution of sources with two
maxima at slightly different radii in the disk [maximum emitted flux is
superposition of two Gaussian curves, each given by equation
(\ref{f2})]. Two neighbouring peaks are determined by the Keplerian
frequencies at radii corresponding to the local maxima in the number
density of the clumps distribution.
\label{fig3}}
\end{fv}

\section{{\mbox{\protect\boldmath{$Q$}}}-Factor and Harmonic
Content of the Power Spectrum}
We assumed modulation of the observed signal by self-radiating or
irradiated clumps orbiting on circular orbits in the Schwarzschild
metric. The superposition of independent contributions from a large
number ($N\gg1$) of such clumps and the noise component are the only
sources of variability in this toy model. A similar situation has been
discussed in various contexts. For example, in the beat-frequency model
of QPOs one considers modulation of the signal by clumps near the sonic
point (Miller et al.\ 1998). However, the assumption about circular
orbits is clearly a simplification which calls for further discussion.
Stella and Vietri (1999) suggest that eccentric orbits of the clumps
could explain non-constant separation of twin kHz peaks detected, e.g.,
in Sco X-1 and 4U 1608$-$52. Each clump represents a spot revolving with
Keplerian frequency $\nu_{\rm{k}}=M^{1/2}r^{-3/2}/(2\pi)$ (geometrized
units). (For illustration: Keplerian frequency in Schwarzschild metric
with $M=2M_{\odot}$ ranges down from its maximum value of 1.1\,kHz which
is acquired at $R_{\rm{ms}}$.) All of these spots reside in a plane
which coincides with an equatorial disk. The propagation of light from a
source near a black hole to a distant observer has been treated by many
authors within different contexts; we used one of the codes developed in
the past to calculate the observed profiles of frequency integrated
light curves within a geometrical-optics approximation (Karas et al.\
1992; see Kato et al.\ 1998 for a recent text-book exposition of the
subject and further references). The orbital radius $R_{\rm{s}}$ and the
observer's inclination $\theta_0$ are two free parameters, which
determine the profile produced by a single source. It turns out that
the size $d$ of the spot and its detailed shape are less important
provided $d{\ll}R_{\rm{s}}$; this condition is assumed hereafter. Given
the two parameters, numerically calculated light curves were
approximated by simple formula which captures the shape of the profile
(the Doppler and the lensing enhancements included). This approach
enables us to carry out fast superposition of hundreds of mutually
independent sources spread over a certain range of $R_{\rm{s}}$ (Karas
1997). To do this, one also adopts that lifetime of the sources is much
longer than their orbital period; one must prescribe the radial
dependence of the local emissivity (e.g., a power-law). The resulting
light curve in the time domain is the sum
$F(\theta_0)=\sum_{n=1}^{N}F_n(R_{\rm{s}},\theta_0)$, and the power
spectrum, $P(\nu)$, is obtained by the Fourier transform
$F{\rightarrow}\hat{F}$ in the standard manner (Born, Wolf 1964):
$2P=\lim_{t\rightarrow\infty}\left[t^{-1}|\hat{F}|\right]$.

The clumps are all placed slightly above the marginally stable orbit,
$R_{\rm{ms}}=3R_{\rm{g}}$ (where gravitational radius in terms of the
central mass is $R_{\rm{g}}=2.95\times10^5M/M_{\odot}$). Their
distribution is characterized by the range of radii ${\Delta}R$ between
$R_{\rm{min}}$ and $R_{\rm{max}}$ ($R_{\rm{min}}{\gta}R_{\rm{ms}}$; see
figure 1), and by the corresponding radial dependence of the maximum flux
from individual clumps,
\begin{equation}
F_{\rm{n,max}}\propto{R_{s}^{-\alpha_1}},
\label{f1}
\end{equation}
where $\alpha_1={\rm{const}}$. We assumed locally isotropic emissivity
of all sources in the frame corotating with the disk matter. The clumps
are generated with random orbital phases and discarded after several
orbits. The fluctuating signal is mean-subtracted and submitted to the
Fourier analysis.

We also explored various modifications of the clumps distribution;
particularly, the power-law (\ref{f1}) (with sharp edges at $R_{\rm{min}}$
and $R_{\rm{max}}$) can be substituted by a smooth Gaussian-type
profile,
\begin{equation}
F_{\rm{n,max}}\propto\exp
\left[-\alpha_{2}\left(R_{\rm{s}}-R_{\rm{cen}}\right)^2\right],
\label{f2}
\end{equation}
where $R_{\rm{cen}}=R_{\rm{min}}+\frac{1}{2}\Delta{R}$,
$\alpha_{2}={\rm{const}}>0$. Such a flux distribution leads to a more
familiar shape of peaks in the power spectrum, but, obviously, the
choice of a particular relation requires one to specify the underlying
physical mechanism producing and maintaining the clumps (which is beyond
the scope of the present paper). It has been proposed and discussed by
several authors that a gap in the disk can be opened where gas is swept
out. This is caused by resonances due to the commensurability of the
orbital period of gas with a distant disturbing secondary, or by the
direct presence of a dense body (a `planet') in the disk, which provokes
the formation of inhomogeneities on the boundary of the gap. The
resulting distribution of spots can then be approximated by superposing
several profiles (\ref{f2}) with different $R_{\rm{cen}}$ and
$\alpha_{2}$'s.

\begin{fv}{4}
{0pc
\epsfxsize=\hsize\epsfbox{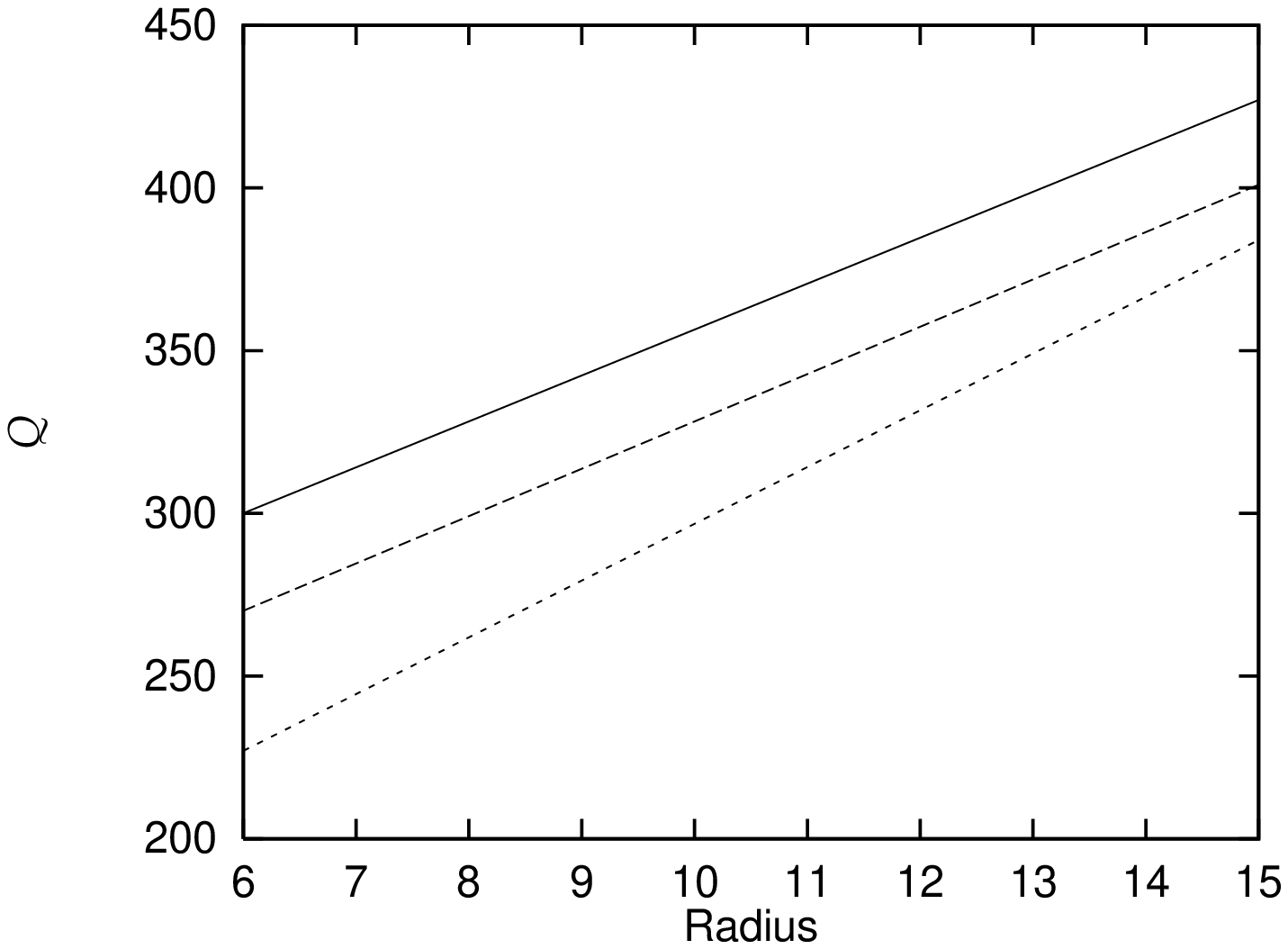}
}
{\protect\small{}Quality factor $Q$ plotted as function of the dimensionless
radius $R_{\rm{min}}/M$ (minimum radial distance of the clumps). This
plot illustrates how $Q$ increases with $R_{\rm{min}}$. The values of $Q$
are less if the clumps are distributed over wider range of radii, as
characterized by $\Delta{R}$. The three lines correspond to
$\Delta{R}=1M$ (solid); $\Delta{R}=2M$ (dashed); $\Delta{R}=3M$
(dotted). $Q$ is only weakly sensitive to other parameters (see the
text).
\label{fig4}}
\end{fv}

Figure 2a shows the power-spectrum density predicted by our simulation.
The graph exhibits a typical peak around the fundamental orbital
frequency of the clumps,
$\nu\approx(3$--$4)\times10^{-3}{\rm{cm}}^{-1}$, and its harmonics. This
plot corresponds to eq.\ (\ref{f2}) with $N=250$, $R_{\rm{min}}=11M$,
$\Delta{R}=3M$, $\theta_0=75^{\rm{o}}$, and $\alpha_{2}=6$. Conversion
to physical units is achieved by multiplying the frequencies in
geometrized units by the factor
$2\times10^5\left(M/M_{\odot}\right)^{-1}$. In figure 2b, in order to
account for high-frequency noise, we also superposed simulated data with
a noise component characterized by zero mean and a standard deviation of
$\sigma=5$\%. The noise is naturally enhanced at higher frequencies, but
does not greatly influence the values of $Q$, which refer to the peaks
in the Fourier spectrum at lower frequency.

Figure 3 illustrates a more complicated situation when the radiation
flux emitted by the clumps in the disk has two maxima at different radii
(superposition of two profiles (\ref{f2}) with $\theta_0=25^{\rm{o}}$
and different values of $R_{\rm{cen,1}}=15M$, $R_{\rm{cen,2}}=17M$;
other parameters as in figure 2). Motivation for such double-peaked
radial distribution arises from possible formation of a gap in the disk.
The fundamental peak in the resulting power spectrum is split into two
neighbouring components, and the signal has a strong harmonic content
even at moderate inclination.

Figure 4 shows the resulting quality factors for different distributions
of the clumps. Here, we give the {\em{maximum expected\/}} $Q$
corresponding to equation (\ref{f1}) with no noise component present in
the signal (the values are decreased by noise, which increases the width
of the peaks). Other parameters of the distribution were $N=250$,
$\alpha_1=1$. Here, the values of $Q$ were averaged over different
inclinations: $\theta_0=25^{\rm{o}}$, $50^{\rm{o}}$, and $75^{\rm{o}}$.
It turns out that $Q$ depends only weakly on $\theta_0$. This conclusion
can be understood by realizing that the redshift factor depends (up to
relativistic corrections; Cunningham 1975) on $\cos\theta_0$ and
determines both the centroid energy of radiation from an orbiting source
and the full width of the observed feature. Consequently, the
$\theta_0$-dependence of $Q$ is not particularly strong provided one
accepts the restriction $\theta_0\lta75^{\rm{o}}$ and
$r{\gta}R_{\rm{ms}}$, as we do in this graph. However, let us note that
the harmonic content of the signal does depend on the inclination. This
is because the light curve of a source orbiting at small inclination
resembles a sine profile, while it becomes distorted by the lensing
effect for large $\theta_0$ ($\gta70^{\rm{o}}$). In other words, the
power in higher harmonics increases when the inclination increases
and/or the mean lifetime of the spots decreases (to be specific, we
assumed 5 orbital periods as typical lifetime in our examples). The
power spectrum thus has potential to reveal orientation of the disk
plane. On the other hand, an excessive amount of power in the overtones
can be used to restrict the validity of the model. The above-given
results are not very sensitive to the assumed local size of the clumps,
provided that it is much less than $R_{\rm{s}}$, and to the life-time of
the spots provided it is much longer than corresponding orbital period.

In our toy model, it appears quite natural that $Q$ increases with the
radius, as can be seen in figure 4 ($\Delta{R}/R_{\rm{min}}$ decreases
when radius increases). One should, however, notice that $Q\gta10^2$
also near $R_{\rm{ms}}$. These results are not very sensitive to the
values of the free parameters, and one can thus consider comparisons
with observational data for particular objects (work in progress).

\section{Conclusions}
We presented a quantitative study of the expected variability power
spectrum relevant for models of quasi-periodic oscillations. The orbital
motion of the clumps and their distribution within a narrow range of
radii are essential. The adopted phenomenological description of the
source employs clumps orbiting in the Schwarzschild metric, and it can be
considered to be the first approximation to more realistic models of
inhomogeneous disk-type accretion flows around black holes and neutron
stars. Although incompleteness of the present treatment results in free
parameters which are not strictly determined here, we checked that
the main results (namely, the quality parameter $Q$) are not strongly
sensitive to reasonable variations of parameter values.

The discussion of the power spectrum presented in this paper is a standard
starting point of a Fourier analysis. It will be extended further by
also examining the Fourier phases and their possible, coherence which should
naturally appear due to the orbital motion of the clumps with the lifetimes
substantially exceeding the dynamical time (Abramowicz et al.\ 1997; Karas
1997). It appears that the wavelet analysis (e.g., Starck et al.\
1998) is better suited for this task than more usual approach
which is offered by Fourier analysis.

\par
\vspace{1pc}
\par
The author thanks for hospitality of the International Center for
Theoretical Physics in Trieste, and for support from the grants GAUK
63/98, GACR 202/\lb{2}98/\lb{2}0522, and GACR 205/\lb{2}97/\lb{2}1165 in
Prague.

\section*{References}
\small
\parskip 1pt
\re
Abramowicz M.~A., Bao G., Lanza A., Zhang X.-H.\ 1991, A\&A 245, 454
\re
Abramowicz M.~A., Bao G., Larsson S., Wiita P.~J.\ 1997, ApJ 489, 819
\re
Bao G., {\O}stgaard E.\ 1995, ApJ 443, 54
\re
Born M., Wolf E.\ 1964, Principles of Optics (Pergamon Press, Oxford),
chapter~1
\re
Cunningham C.~T. 1975, ApJ 202, 788
\re
Karas V.\ 1997, MNRAS 288, 12
\re
Karas V., Vokrouhlick\'y D., Polnarev A.\ 1992, MNRAS 259, 569
\re
Kato S., Fukue J.\ 1980, PASJ 32, 377
\re
Kato S., Fukue J., Mineshige S.\ 1998,
Black-Hole Accretion Disks (Kyoto University Press, Kyoto),
chapter~3
\re
Markovi\'c D., Lamb F.~K.\ 1998, ApJ 507, 316
\re
Matt G., Perola G.~C., Piro L.\ 1991, A\&A 247, 25
\re
Miller M.~C., Lamb F.~K., Psaltis D.\ 1998, ApJ 508, 791
\re
Papadakis I.~E., Lawrence A.\ 1995, MNRAS 272, 161
\re
Pariev V.~I., Bromley B.~C.\ 1998, ApJ 508, 590
\re
Stella L., Vietri M.\ 1999, Phys.\ Rev.\ Lett.\ 82, 17
\re
Starck J.-L., Murtagh F., Bijaoui~A.\ 1998, Image Processing and
Data Analysis (Cambridge University Press, Cambridge), chapter~1
\re
van der Klis M.\ 1999, in Proceedings of the Third William Fairbank
Meeting, in press; astro-ph/9812395

\label{last}
\end{document}